\documentclass{ametsocV6.1}
\usepackage{multirow}
\usepackage{hyperref}
\makeatletter
\newcommand\notsotiny{\@setfontsize\notsotiny\@vipt\@viipt}
\makeatother

\title{Prediction of Tropical Pacific Rain Rates with Over-parameterized Neural Networks}

\authors{Hojun You \aff{a},\thanks{This Work has not yet been peer-reviewed and is provided by the contributing Author(s) as a means to ensure timely dissemination of scholarly and technical Work on a noncommercial basis. Copyright and all rights therein are maintained by the Author(s) or by other copyright owners. It is understood that all persons copying this information will adhere to the terms and constraints invoked by each Author's copyright. This Work may not be reposted without explicit permission of the copyright owner.}
Jiayi Wang \aff{b},
Raymond K.~W.~Wong\aff{c} ,
Courtney Schumacher\aff{d}, 
R. Saravanan\aff{d},
and Mikyoung Jun\aff{a} 
\correspondingauthor{Mikyoung Jun, mjun@central.uh.edu} 
}

\affiliation{\aff{a}{Department of Mathematics, University of Houston, Houston, TX, USA}\\
\aff{b}{Department of Mathematical Sciences, University of Texas at Dallas, Richardson, TX, USA}\\
\aff{c}{Department of Statistics, Texas A\&M University, College Station, TX, USA}\\
\aff{d}{Department of Atmospheric Sciences, Texas A\&M University, College Station, TX, USA}
}

\abstract{The prediction of tropical rain rates from atmospheric profiles poses significant challenges, mainly due to the heavy-tailed distribution exhibited by tropical rainfall. This study introduces over-parameterized neural networks not only to forecast  tropical rain rates, but also to explain their heavy-tailed distribution. The prediction is separately conducted for three rain types (stratiform, deep convective, and shallow convective) observed by the Global Precipitation Measurement satellite radar over the West and East Pacific regions. Atmospheric profiles of humidity, temperature, and zonal and meridional winds from the MERRA-2 reanalysis are considered as features. Although over-parameterized neural networks are well-known for their ``double descent phenomenon," little has been explored about their applicability to climate data and capability of capturing the tail behavior of data. In our results, over-parameterized neural networks accurately predict the rain rate distributions and outperform other machine learning methods. Spatial maps show that over-parameterized neural networks also successfully describe spatial patterns of each rain type across the tropical Pacific. In addition, we assess the feature importance for each over-parameterized neural network to provide insight into the key factors driving the predictions, with low-level humidity and temperature variables being the overall most important. These findings highlight the capability of over-parameterized neural networks in predicting the distribution of the rain rate and explaining extreme values.} 

\begin{document}

%% Necessary!
\maketitle

%%%%%%%%%%%%%%%%%%%%%%%%%%%%%%%%%%%%%%%%%%%%%%%%%%%%%%%%%%%%%%%%%%%%%
% SIGNIFICANCE STATEMENT/CAPSULE SUMMARY
%%%%%%%%%%%%%%%%%%%%%%%%%%%%%%%%%%%%%%%%%%%%%%%%%%%%%%%%%%%%%%%%%%%%%
%
% If you are including an optional significance statement for a journal article or a required capsule summary for BAMS 
% (see www.ametsoc.org/ams/index.cfm/publications/authors/journal-and-bams-authors/formatting-and-manuscript-components for details), 
% please apply the necessary command as shown below:
%
% Significance Statement (all journals except BAMS)
%
\statement
    This study aims to introduce the capability of over-parameterized neural networks, a type of neural network with more parameters than data points, in predicting the distribution of tropical rain rates from grid-scale environmental variables and explaining their tail behavior. Rainfall prediction has been a topic of importance, yet it remains a challenging problem for its heavy-tailed nature. Over-parameterized neural networks correctly captured rain rate distributions and the spatial patterns and heterogeneity of the observed rain rates for multiple rain types, which could not be achieved by any other previous statistical or machine learning frameworks. We find that over-parameterized neural networks can play a key role in general prediction tasks, with potential expanded applicability to other domains with heavy-tail data distribution.
%
%% Capsule (BAMS only)
%%
%\capsule
%       Enter BAMS capsule here, no more than 30 words. See \url{www.ametsoc.org/index.cfm/ams/publications/author-information/formatting-and-manuscript-components/#capsule} for details.
% 
%% * * If using twocol mode, you will need to use the commands "twocolsig" and "twocolcapsule" in place of "sig" and "capsule"
%%      to ensure that the text box correctly spans across both columns.

%%%%%%%%%%%%%%%%%%%%%%%%%%%%%%%%%%%%%%%%%%%%%%%%%%%%%%%%%%%%%%%%%%%%%
% MAIN BODY OF PAPER
%%%%%%%%%%%%%%%%%%%%%%%%%%%%%%%%%%%%%%%%%%%%%%%%%%%%%%%%%%%%%%%%%%%%%
%
\section{Introduction}

The tropical Pacific region plays a crucial role in global climate variability. Rainfall anomalies associated with the El Ni\~{n}o-Southern Oscillation phenomenon generate atmospheric waves that propagate across the globe and impact weather in remote regions of the globe. However, climate models still suffer from biases in this region that affect assessments of weather and climate risk \citep{lee2022biases, sobel2023biases}. Several attempts have been made to use statistical and machine learning frameworks to predict rain rates in the tropics. Previous attempts applied machine learning techniques to the parameterization of convection using cloud-resolving model or high-resolution model simulations \citep{brenowitz2018prognostic, o2018using, rasp2018deep}. These efforts bring computational gains, but they cannot overcome inherent model deficiencies and may not identify the underlying physical relationships. Instead, \citet{yang2019predictive} sought to capture the relationship between atmospheric features and the rain amount by applying a generalized linear model (GLM) to observational data. GLMs naturally provide the relationship through their coefficients. But, they may struggle to predict the tail of the rain rate distributions accurately due to their parametric assumption on the density function. Building upon this work, \citet{wang2021statistical} improved the predictive performance by implementing random forest (RF) and neural network (NN) methods. These machine learning methods allowed for more flexibility of the relationship and demonstrated better performance than GLM in predicting the rain rate distribution and capturing its tail behavior. However, the distribution from their outputs still deviated significantly from the observations. In this work, we experiment with over-parameterized NNs, where the number of NN parameters exceed the sample size, and find that they enhance the previous results and effectively capture the tail of the rain rate distribution.

\begin{figure}[!ht]
    \centering
    \includegraphics[width=.8\linewidth]{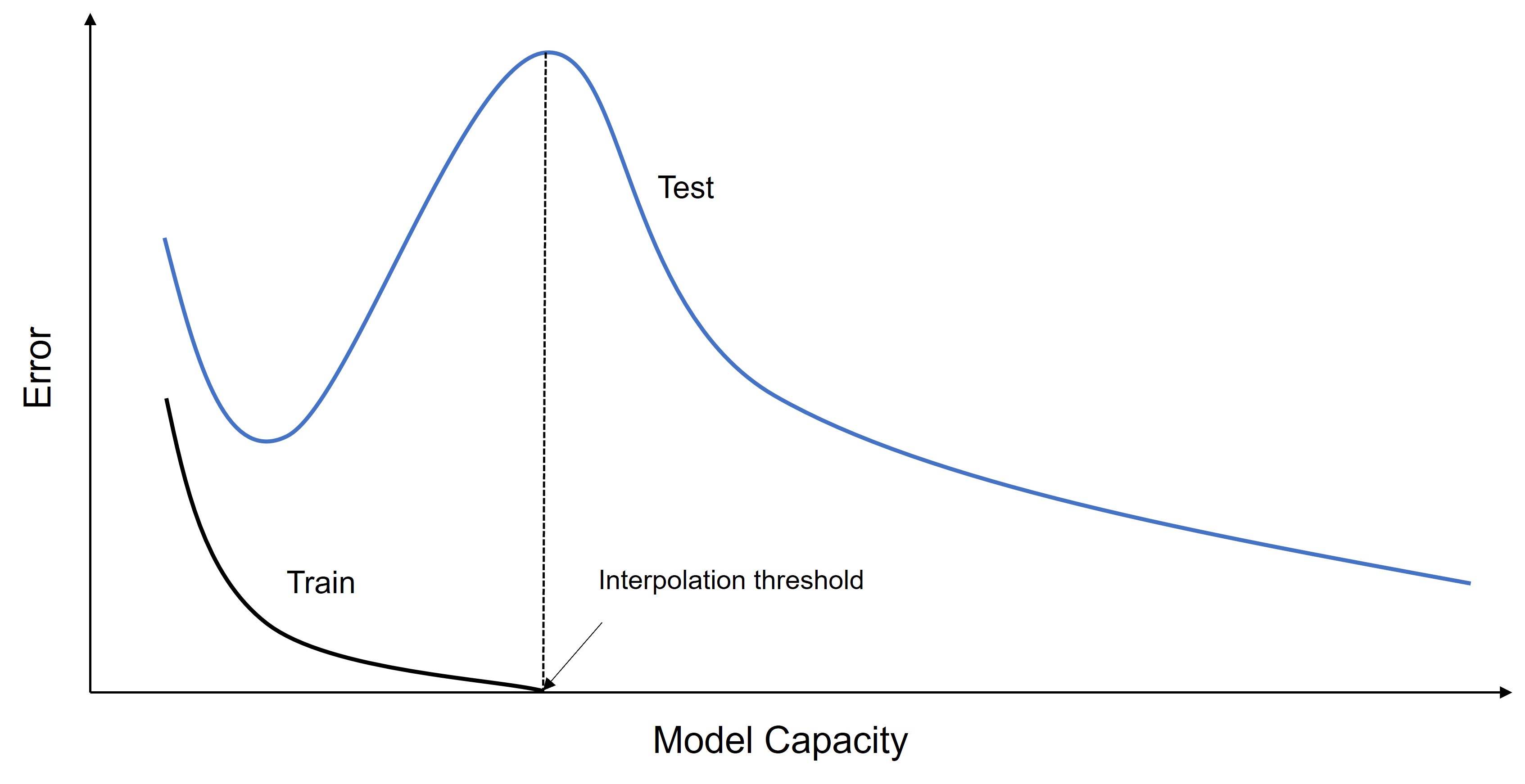}
    \caption{Illustration of the double descent phenomenon. The training curve is represented by the black line, while the test curve is depicted in blue. The point on the x-axis where the training error reaches zero is ``interpolation threshold."}
    \label{fig:doubledescent}
\end{figure}

The over-parameterized regime has gained significant attention since the \textit{double descent} phenomenon was proposed by \citet{belkin2019reconciling}. The double descent describes the pattern of the test error curve with respect to the model capacity (Figure \ref{fig:doubledescent}). According to classical wisdom, the test curve exhibits a U-shaped pattern before reaching the interpolation threshold, where perfect interpolation can be attained. \citet{belkin2019reconciling} showed that beyond the interpolation threshold, the test error decreases again, creating a double descent pattern of the test curve. Interestingly, the notorious issue of overfitting, where an excessive number of model parameters compared to the number of samples harms the test performance, does not show up beyond the interpolation threshold. Since \citet{belkin2019reconciling}, numerous studies have been conducted to explore and explain this phenomenon theoretically and experimentally \citep{geiger2019jamming,bartlett2020benign, d2020double, nakkiran2020optimal,   nakkiran2021deep, muthukumar2021classification,  gamba2022deep,hastie2022surprises}. However, most of these studies focused on classification tasks, with little attention given to leveraging the double descent phenomenon for accurate regression predictions on real datasets. In addition, to the best of our knowledge, the capability of over-parameterized NNs in explaining the tail behavior of data has never been investigated. In this study, we introduce over-parameterized NNs to rainfall predictions and present numerical findings that demonstrate the efficacy of this approach.

% Another crucial aspect of interest in the rainfall dataset is identifying key features for our model. However, a challenge with machine learning methods such as NNs is their lack of interpretability. NNs capture complex relationships between variables, making it difficult to disentangle the individual contributions of each feature. To address this issue, we employed the SHapley Additive exPlanations (SHAP) value approach devised by \cite{lundberg2017unified}. The SHAP values allow us to assess the feature importance by examining the change in the expected model output conditioned on a specific feature. By adopting the SHAP value approach, we were able to provide insights into the key features for our over-parameterized NNs and enhance the interpretability of our model.
Another crucial aspect of interest in the rainfall dataset is identifying key features for our model. Even if NNs successfully capture complex relationships between variables, it may be difficult to disentangle the individual contributions of each feature. To address this issue, we employed the permutation importance (PI) approach devised by \cite{breiman2001random}. The PI approach allows us to assess the feature importance by examining the change in the model output with permuted feature values from the original model output. By adopting the PI approach, we are able to provide insights into the key features for our over-parameterized NNs and enhance the interpretability of our model.

The structure of the remainder of this paper is as follows. We provide a detailed description of the data used in our study in Section 2. In Section 3, we recap machine learning methods used in previous studies and introduce the over-parameterized regime with a suitable training method. Section 4 first checks if our over-parameterized NNs are properly trained and exhibit the double descent phenomenon. Then, we compare prediction results of the rain rate distribution with observations and other machine learning methods. Lastly, we summarize our findings and discuss future work.

\section{Data description}

\begin{figure}[!ht]
    \centering
    \includegraphics[width=\textwidth]{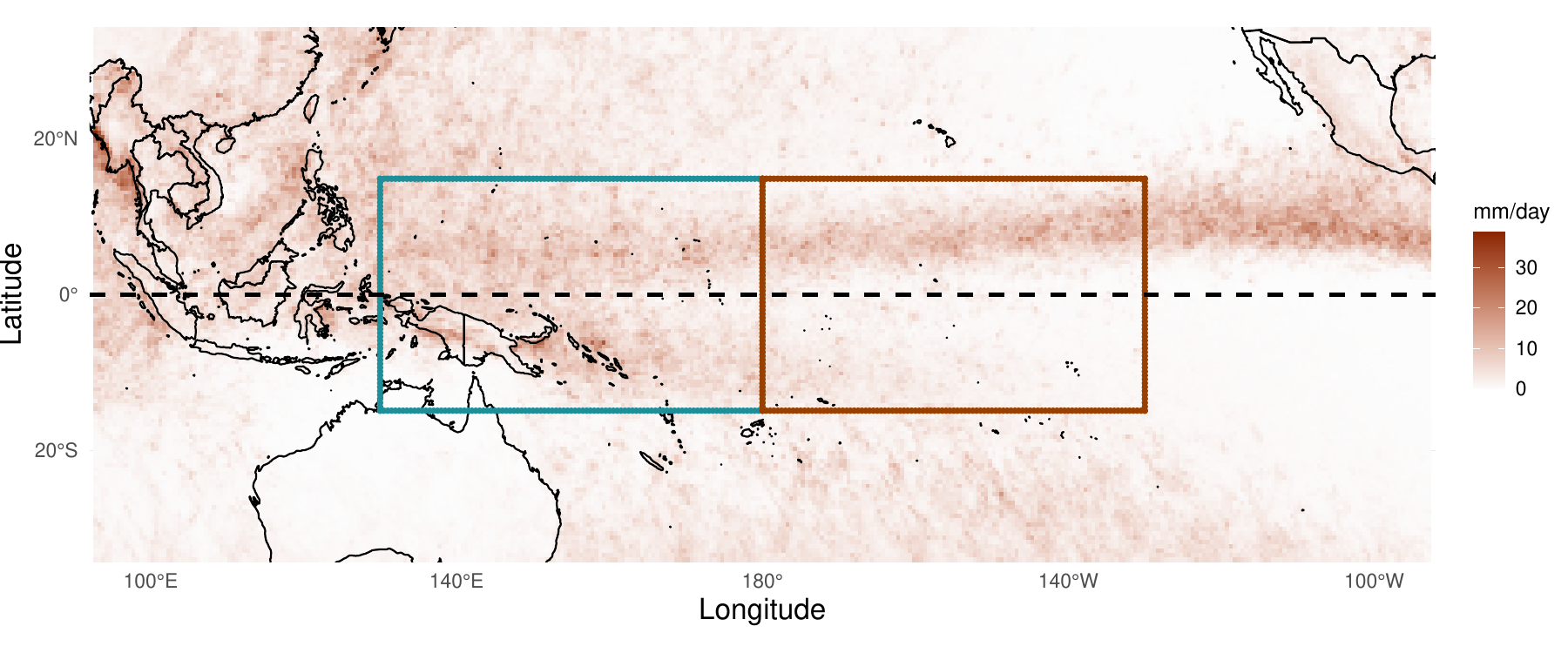}
    \caption{The average GPM DPR rain rate (in mm/day) during JJA 2015-2022 over the tropical Pacific. The data domains for this study were separated into the West Pacific (blue) and East Pacific (red) regions.}
    \label{fig:domain}
\end{figure}

We used eight years of June, July and August (JJA) rain rate data from 2015 to 2022 (Figure \ref{fig:domain}). The rain rate observations were obtained from the Global Precipitation Measurement (GPM) dual-frequency precipitation radar (DPR) Version 7 dataset \citep{iguchi2021gpm}. JJA data from 2015-2018 were used for training and JJA data from 2019-2022 were used for testing. The observational domain was limited to the tropical West Pacific (WP; $130^{\circ}\text{E}-180^{\circ}\text{E}$, $15^\circ \text{S}-15^{\circ}\text{N}$) and East Pacific (EP; $180^{\circ}\text{W}-130^{\circ}\text{W}$, $15^\circ \text{S}-15^{\circ}\text{N}$) regions, as illustrated in Figure \ref{fig:domain}. These two regions were selected for their distinct convective environments, with the WP box representing the warm pool region and South Pacific convergence zone (SPCZ) and the EP box impacted by equatorial upwelling with a more distinct intertropical convergence zone (ITCZ) north of the equator. The EP box is also where global climate models (GCMs) commonly experience an erroneous double ITCZ in boreal summer \citep[e.g.,][]{oueslati2015double}. 

The orbital DPR data was gridded to a temporal resolution of 3 hours and a spatial resolution of $0.5^{\circ}$, resulting in 6,000 spatial locations for each region. We chose this time and space resolution to match the specifications of higher-resolution GCM output; however, it is important to note that the DPR data represents just snapshots within the 3-hour period because of its scanning geometry. GPM is in an inclined low-earth orbit, so it only revisits particular points on the Earth's surface every day or so, and the times of day vary because of the precessing nature of the orbit (although this feature of the orbit is important to capture the full diurnal cycle). The DPR has a footprint size of 5 km at nadir and a swath width of 245 km that samples only part of each domain each day. To ensure robust sampling in each grid box, we only consider $0.5^{\circ}$ grids that had an overpass containing at least 50 DPR pixels (or about half of the grid box) regardless of whether the pixels were raining or not. We average the mean DPR-observed near-surface rain rate over the entire $0.5^{\circ}$ grid for a particular 3-hour period and a particular rain type. Therefore, the grid-averaged values will be lower than instantaneous 5-km pixels values.

For the rain types, we adopted a three-type categorization: stratiform, deep convective, and shallow convective \citep{funk2013analysis}. This separation allows us to account for the unique nature of each rain type, with deep convection extending through the depth of the troposphere due to its stronger updrafts and shallow convection being confined to echo tops below the $0^{\circ}$C level because of weaker updrafts and/or a limiting environment (such as dry air) at upper levels \citep{schumacher2023assessing}. Stratiform rain in the tropics forms from deep convection and is an important component of mesoscale convective systems \citep{houze1997stratiform}. The percent of total rain for each rain type in each domain is summarized in Table \ref{tab:data_summary}. As our focus is on predicting the rain rate, we only use samples with non-zero rain rates, with numbers of samples for each region indicated in Table \ref{tab:samples}. Tables \ref{tab:data_summary} and \ref{tab:samples} show that while the deep convective rain type has the lowest frequencies in both the WP and EP regions, it contributes about $40\%$ of the total rain. On the contrary, the shallow convective rain type shows the highest frequencies in both regions, yet it makes up the smallest portion of the total rain amount  (about 10-20\%) in both regions. This emphasizes the distinct nature of each rain type and demonstrates the importance of separately considering each rain type.

\begin{table}[!ht]
    \centering
    \caption{Rain amount percentage of each rain type to the total rain amount in the West Pacific and East Pacific regions during JJA.}
    \begin{tabular}{ccc}
    \topline
        Rain type & WP & EP \\ \midline
        Stratiform & 47.2 & 42.1 \\
        Deep convective & 42.9 & 37.7\\
        Shallow convective & 9.9 & 20.2\\\botline
    \end{tabular}
    \label{tab:data_summary}
\end{table}

\begin{table}[!ht]
    \centering
    \caption{The number of $0.5^{\circ}$ grid boxes with positive rain rate for each rain type and region in the training (JJA 2015-2018) and testing (JJA 2019-2022) datasets.}
    \begin{tabular}{ccccc}
    \topline
        \multirow{2}{*}{Rain type} & \multicolumn{2}{c}{WP} & \multicolumn{2}{c}{EP} \\ \cline{2-5}
         & Train & Test & Train & Test \\ \midline
         Stratiform & 160,170 & 157,682 & 94,034 & 72,898 \\
         Deep convective & 113,719 & 115,052 & 67,735 & 52,468 \\
         Shallow convective & 207,186 & 208,982 & 185,353 & 170,386 \\ \botline
    \end{tabular}
    \label{tab:samples}
\end{table}

As in \citet{yang2019predictive}, we consider humidity, temperature, zonal wind, meridional wind, latitude, longitude, and latent heat flux as features for the rain rate prediction. The features used in this study were computed from the MERRA-2 reanalysis \citep{rienecker2011merra}. The MERRA-2 humidity, temperature, zonal wind, and meridional wind fields consist of vertical profiles at 40 pressure levels, leading to 163 features in total including latitude, longitude and latent heat flux.  MERRA-2 data are available at the same temporal and spatial resolutions (3 hourly and $0.5^{\circ}$) as the DPR rain rate dataset. The MERRA-2 grid matching the time and location of the DPR overpass was used in the training process. MERRA-2 grids not matching DPR overpasses were ignored.

Rainfall prediction presents a notable challenge due to the extremely thick tail in the rain rate distribution, indicating a relatively large portion of extreme values. Table \ref{tab:heavy} summarizes some statistics of the training dataset for each rain type and region. The percentiles in Table \ref{tab:heavy} represent the ranking of a particular value within a dataset. For instance, the 75th percentile indicates that 25\% of the data values are above it and the 99th percentile means that only 1\% of the data values exceed it. In the case of stratiform and deep convective rain, the mean exceeds both the median and the 75\% percentile value. This indicates that the data is highly heavy-tailed, as the mean values are significantly influenced by the presence of extreme values. Although shallow convective rain may not be as heavy-tailed as the other two rain types, the mean values lying between the median and 75\% percentile values still points out the presence of a thick tail in the data. 

\begin{table}[!ht]
    \centering
    \caption{Mean, percentiles (median, $75\%$, $90\%$, $99\%$), and maximum DPR rain rate value (in mm/hr) of the training dataset for each rain type in each region for a $0.5^{\circ}$ grid resolution.}
    \begin{tabular}{ccrrrrrr}\topline
        Region & Rain type & Mean & Median & 75\% & 95\% & 99\% & Max \\ \midline
        \multirow{3}{*}{WP} & Stratiform & 0.42 & 0.10 & 0.31 & 1.86 & 5.42 & 28.74\\
         & Deep convective & 0.48 & 0.08 & 0.31 & 2.23 & 6.57 & 60.00\\
         & Shallow convective & 0.05 & 0.03 & 0.07 & 0.13 & 0.32 & 2.35 \\ \midline
        \multirow{3}{*}{EP} & Stratiform & 0.34 & 0.07 & 0.24 & 1.48 & 4.79 & 30.30\\
         & Deep convective & 0.47 & 0.10 & 0.34 & 2.15 & 6.18 & 50.38\\
         & Shallow convective & 0.08 & 0.04 & 0.10 & 0.19 & 0.48 & 2.97 \\ \botline
    \end{tabular}
    \label{tab:heavy}
\end{table}

A challenge from dealing with heavy-tailed data arises from a large portion of extreme samples. These extreme values can distort the overall patterns from the majority of the data and, hence, decrease the predictive power of models. In addition, many traditional methods may lack the necessary flexibility to cover the full range of values from a majority of the data to the extreme values. As a result, these methods may struggle to estimate and predict the data, resulting in unsatisfactory overall performance. This limitation can particularly affect a model's ability to accurately capture and forecast extreme events, which is crucial in tasks like rainfall forecasting. 

\section{Methods}

In the following, we present four statistical and machine learning frameworks to investigate the relationship between the rain rates and features. 

\subsection{Generalized linear model}

The GLM \citep{mccullagh1989generalized} is a flexible statistical model that provides an extension to ordinary linear regression by allowing non-normal distribution of the response variable. In contrast to ordinary linear regression, where the mean of the response, $\bm\mu$, is directly modeled as a linear function of the features, the GLM introduces the link function to establish a nonlinear relationship between the response and the features. The link function enables the GLM to handle a wider range of data distributions and provide greater capability in modeling complex relationships. 

The GLM consists of three essential components: the random component, the systematic component, and the link function. The random component deals with the distributional assumption of the response variable, allowing for non-normal distributions. The systematic component represents the linear predictor, which is formed by combining the features with their regression coefficients. The link function constructs a relationship between the random and systematic components, i.e.,
$$E(Y | \bm X) = \bm \mu = g^{-1}(\bm X^{\top} \bm\beta),$$
where $Y \in \mathbb{R}$ is the response, $\bm X \in \mathbb{R}^p$ is the feature with $p$ dimensions, $\bm\beta \in \mathbb{R}^p$ is the coefficient of $\bm X$, and $g$ is the link function. Depending on the distribution of the response and the link function, GLMs can take on different names such as logistic regression, Gamma regression, etc. 

Following \citet{yang2019predictive} and \citet{wang2021statistical}, we implement Gamma regression for its resemblance to the distribution of rainfall in terms of skewness. Suppose $Y(\bm s)$ is a response at location $\bm s \in \mathcal{S} \subset\mathbb{R}^d$, where $\mathcal{S}$ is an observation domain. Then, Gamma regression, with the log-link function $g(x)=\log(x)$, has the following structure: 
\begin{align*}
    & Y(\bm s) \sim Gamma(\alpha(\bm s), \beta(\bm s)), \\
    & E(Y(\bm s) | \bm X(\bm s)) = \alpha(\bm s)\cdot \beta(\bm s) = \mu(\bm s), \\
    & \log(\mu(\bm s)) = \eta_0+\eta_1 X_1(\bm s)+\cdots \eta_p X_p(\bm s),
\end{align*}
where $\alpha(s)>0$ and $\beta(s)>0$ are parameters from the Gamma distribution where a probability density function is given as $$f(y;\alpha, \beta)=\displaystyle\frac{1}{\beta^{\alpha}\Gamma(\alpha)}y^{\alpha-1} e^{-\frac{y}{\beta}},~~ 0<y<\infty.$$ Here, $\Gamma$ is a Gamma function, $\bm X(\bm s)$ is the feature vector at location $\bm s$, and $\eta_i\text{'s}\ (i=0, \ldots, p)$ are coefficients for each feature with the total number of $p$. Parameter estimation in Gamma regression can be done by maximum likelihood estimation%iteratively reweighted least squares
, which is provided by \texttt{R} with \texttt{glm} function.

Despite some advantages of Gamma regression, such as its ability to handle skewed data, its limited parametric distribution we assume for the error may not adequately capture the characteristics of heavy-tailed data such as rainfall  \citep{yang2019predictive, wang2021statistical}. Thus, adopting machine learning techniques, which do not impose such strong restrictions as GLMs, could potentially enhance the prediction performance.

\subsection{Random forest}

RF has gained attention for its powerful prediction capability since it first appeared in \citet{breiman2001random}. RF is an ensemble method that aggregates multiple decision trees to achieve accurate and robust predictions. A decision tree is a non-parametric machine learning method that has a hierarchical tree structure. It consists of three key components: internal nodes, branches, and leaf nodes. Each internal node represents a condition on one of the features. The branches indicate split data based on the outcomes of the condition at each internal node. The leaf nodes denote the final outcomes from split data at corresponding branches. %Imagine a simple decision tree with just one internal node. The condition at the internal node can be whether the first feature value is positive. Based on this condition, the tree splits the data into two branches: one for samples with positive values of the first feature and the other for samples with negative values of first feature. At last, each leaf node at the end of these branches corresponds to predicted values from the data samples of the branches. 

Ensemble methods summarize the results from multiple statistical or machine learning methods into one final outcome. In an RF model, each decision tree independently learns from random samples of training data with a random subset of features at each split. This random sampling and subsetting process help to bring diversity and reduce the risk of overfitting by preventing the trees from becoming too specialized to the training data or correlated to one another. Aggregating the predictions from multiple less correlated trees results in outcomes with lower variance and improved accuracy.

Although RF may lack interpretability compared to GLM, its performance in fitting and predicting rainfall data exceeded GLM and matched NN \citep{wang2021statistical}. In our study, we train a random forest model in \texttt{R} with \texttt{randomForest} function supported by \texttt{randomForest} package. The number of trees was set to 100 and the number of features selected at each split was set to the largest number below $\sqrt{\text{the total number of features}}$, which is 12. For the other configurations, we used the default values provided with \texttt{randomForest} function.
%\begin{figure}[!hb]
%    \centering
%    \includegraphics[width=\linewidth]{mlp.pdf}
%    \caption{Structure of multi-layer perceptron with 5 layers. The input value is fed into the input layer and the output value comes out from the output layer. Hidden layers are layers that connect from the input layer to the output layer. Every unit of one layer is connected to every unit of the next layer.}
%    \label{fig:mlp}
%\end{figure}
\subsection{Neural networks}

\subsubsection {Multi-layer perceptron NN}

As the size of datasets grow and high-performing computation becomes more available, NNs have been one of the most popular frameworks in the statistical and machine learning communities. Inspired by the structure of the human brain, NNs are computational models which are defined by hidden units and connections between the hidden units. Hidden units and connections correspond to the neurons and the synapses in human brain, respectively. A typical example would be multi-layer perceptron (MLP), which connects all the units from the input to the output \citep{lecun2015deep}. Besides MLP, convolutional NN \citep{simonyan2014very}, recurrent NN \citep{hochreiter1997long}, and transformer \citep{vaswani2017attention} are frequently used structures. In this study, MLP is used to compare standard NN performance with GLM and RF, as our primary goal lies in introducing over-parameterized NNs. The structure of MLP is described below.

% MLP is usually established with an input layer followed by sequential hidden layers and an output layer at the end. %(Figure \ref{fig:mlp}). 
% The number of layers is often called the depth of the NN and the number of units in each layer is called the width of the NN. To take into account nonlinear structure, nonlinear activation functions are applied after each hidden layer. Some popular nonlinear activation functions are Tangent hyperbolic (Tanh), Rectified Linear Unit (ReLU), Gaussian Error Linear Unit (GELU), etc. The choice of the nonlinear activation function often plays a key role in the model performance and optimization. In this study, we chose either the Tanh or ReLU function that best explains each rain type in terms of test errors. At the output layer, an exponential function is applied to guarantee positive values of the outputs. In short, MLP with $L$ layers can be put down in the following equations:
\begin{align*}
\text{Input layer: } & \bm{Z}^{(1)} = \begin{bmatrix} Z_1 \\ Z_2 \\ \vdots \\ Z_p \end{bmatrix} \\
\text{Hidden layer 1: } & \bm{H}^{(1)} = \bm{W}^{(1)}\bm{Z}^{(1)} + \bm{b}^{(1)} \\
& \bm{Z}^{(2)} = \sigma^{(1)}(\bm{H}^{(1)}) \\
\text{Hidden layer 2: } & \bm{H}^{(2)} = \bm{W}^{(2)}\bm Z^{(2)} + \bm{b}^{(2)} \\
& \bm{Z}^{(3)} = \sigma^{(2)}(\bm{H}^{(2)}) \\
& \vdots \\
\text{Output layer: } & \bm{H}^{(L-1)} = \bm{W}^{(L-1)}\bm{Z}^{(L-1)} + \bm{b}^{(L-1)} \\
& Y = \sigma^{(L-1)}(\bm{H}^{(L-1)}),
\end{align*}
where $\bm Z^{(l)}$ is the input vectors to $l$-th layer, $\bm H^{(l)}$ is the hidden outputs from $l$-th linear layer and $Y$ is the final output. $\bm W^{(l)}$'s and $\bm b^{(l)}$'s are the weights and biases of the NNs and $\sigma^{(l)}$ is the activation function at $l$-th hidden layer. 

\subsubsection{Over-parameterized regime}

An over-parameterized regime refers to when the number of parameters in a model exceeds the training sample size. Over-parameterized models have gained significant attention in recent years \citep{cao2022benign, hastie2022surprises, liu2022loss, xu2022infinitely, gao2023over}, particularly with the rise of deep learning, as many state-of-the-art deep learning methods fall into this category. From classical wisdom, an excess number of parameters compared to the number of data points usually leads to overfitting. However, it has been observed that over-parameterized NN models do not suffer from traditional overfitting and result in decent generalization.

\citet{belkin2019reconciling} experimentally showed how classical understanding of overfitting may have to change in an over-parameterized regime. Their study demonstrated that over-parameterized models not only perfectly interpolate the training data, but also attain lower test error than what can be achieved in under-parameterized regimes. According to the classical understanding, the test error curve first decreases until an unknown ``sweet spot" where the minimum test error is obtained, then it starts to increase again. The maximum test error occurs around the interpolation threshold, where the number of parameters matches the sample size. Beyond the interpolation threshold, also called the over-parameterized regime, the test error starts to decrease again. This phenomenon is often called \textit{double descent}. %\citep{belkin2019reconciling}. 

However, it becomes more challenging to train over-parameterized models as we need to explore a larger parameter space. \citet{belkin2019reconciling} suggested a sequential training procedure to gradually increase the model size, with the final weight of smaller models serving as the pre-trained initialization for the larger models. We followed a similar sequential training procedure and considered NNs with different depths ($L=4, 5, 6$) and widths ($W=12, 24, \ldots, 600$). For example, when training NNs with $L=4$, the learned weights of a NN with $L=4$ and $W=12$ were used as the initialization for a NN with $L=4$ and $W=24$. The remaining weights were initialized by default in \texttt{PyTorch}, which in our case was the Kaiming uniform initialization \citep{he2015delving}. For deeper NNs ($L=5, 6$), the same width was maintained, but NNs with one less layer were used for initialization. Our final models were NNs with $L=6$, but NNs with $L=5$ were considered for initialization of NNs with $L=6$, as we found this process stabilized the training procedure. This pre-training strategy allowed us to optimize over a larger parameter space while benefiting from the knowledge gained by smaller models in the sequential progression. 

Every experiment with NNs was conducted by \text{PyTorch} with the Adam optimizer \citep{kingma2014adam}. The total number of epochs for every NN with a different size was set to be 2000 and the learning rate decayed by 0.9 every 100 epochs. To quantify the discrepancy between predicted and target values, we used the mean squared error (MSE) as the loss function. It is important to note that no explicit or implicit regularization techniques, such as dropout or imposing a penalty on weights, were utilized. More details on the training configurations for each region and rain type are listed in Table \ref{tab:nn_config}.

\begin{table}[!ht]
    \centering
    \caption{Training configurations for NNs with 6 layers. WP and EP stand for the West Pacific and East Pacific region, respectively.}
    \begin{tabular}{cccc}\topline
        Region & Rain type & Activation & Initial learning rate \\ \midline
        \multirow{3}{*}{WP} & Stratiform & Tanh & $5 \times 10^{-5}$ \\ 
         & Deep convective & Tanh & $9 \times 10^{-5}$ \\ 
         & Shallow convective & Tanh & $10 \times 10^{-5}$ \\ \midline
        \multirow{3}{*}{EP} & Stratiform & Tanh & $10 \times 10^{-5}$ \\ 
         & Deep convective & ReLU & $1 \times 10^{-5}$ \\ 
         & Shallow convective & Tanh & $10 \times 10^{-5}$ \\ \botline
    \end{tabular}
    \label{tab:nn_config}
\end{table}

Each model was trained on a single Tesla V100 GPU, with training times influenced by both rain type and region due to varying sample sizes. The most data-intensive case, shallow convective rain in the WP region, demanded the longest training due to its largest sample size. In this case, training the smallest model ($L=4,\ W=12$) took 56 minutes, while the largest ($L=6,\ W=600$) required 65 minutes. Due to sequential training procedure, the entire training time for the largest model extended to 55 hours as it requires training of smaller models for initialization. Conversely, the deep convective rain type in the EP region had shorter training times, with the smallest and largest models taking 25 and 26 minutes, respectively. The entire training time for this case was 23 hours.

\subsection{Model interpretation}

One of the popular approaches to explain variable importance in machine learning is permutation importance \citep{breiman2001random, fisher2019all, sood2022feature, cheung2022hybrid, ramos2023using, sekiyama2023surrogate}. Permutation importance quantifies the importance of individual features by assessing the impact on test errors when their values are randomly shuffled while keeping the other features unchanged. To calculate the permutation importance for a specific feature (e.g., temperature at a particular height), we first shuffle values of the target feature while leaving the remaining features in place. Then, we obtain model predictions from this permuted data and compute the increase in the loss values. A larger increase in loss values indicates that the target feature plays a more significant role in the model, as breaking the relationship between this target feature and the model outcome leads to a greater deterioration in model performance. 

The original idea, proposed by \cite{breiman2001random}, aimed to explore local variable importance, i.e., the variable importance for an individual sample. However, our primary objective is to achieve transparency for the entire model, hence we employ permutation importance to derive global variable importance \citep{konig2021relative, sood2022feature}. The only distinction between calculating local and global variable importance lies in whether each permutation is kept distinct or identical across samples for a given target feature. While there are other popular interpretation methods such as SHapley Additive exPlanations (SHAP, \cite{lundberg2017unified}) and Local Interpretable Model-agnostic Explanations (LIME, \cite{ribeiro2016should}), they mainly provide local variable importance. Since our focus is on global variable importance, permutation importance was considered the most suitable choice for this study.

To summarize our variable importance findings, we employed the following process. For each feature, we conducted 100 random permutations, ensuring the robustness of our variable importance investigations. In each permutation, we identified the top 10 most significant features and aggregated these selections across all permutations. Finally, we determined key features for our model by reporting the top 10 most frequently occurring ones within this combined set of features. 

\section{Results}

Only data points with non-zero rain rates were used for the results presented below. We compare four statistical and machine learning methods: GLM, RF, the under-parameterized NN, and the over-parameterized NN. We chose the under-parameterized NN model that yielded the best test error before peaks in the test curve and the over-parameterized NN model with the largest number of parameters. The best models were separately obtained for the WP and EP regions and three different rain types: stratiform, deep convective, and shallow convective. 

\subsection{Double descent}

We first examine whether over-parameterized NNs are well trained. To properly train over-parameterized NNs, the model size should be increased beyond the interpolation threshold, where we can achieve zero training error and observe the double descent curve \citep[Figure 1]{belkin2019reconciling}. Then, over-parameterized models at the end of the double descent curve are used to see how these models perform with heavy-tailed rainfall data. 

\begin{figure}[!ht]
    \centering
    \includegraphics[width=\textwidth]{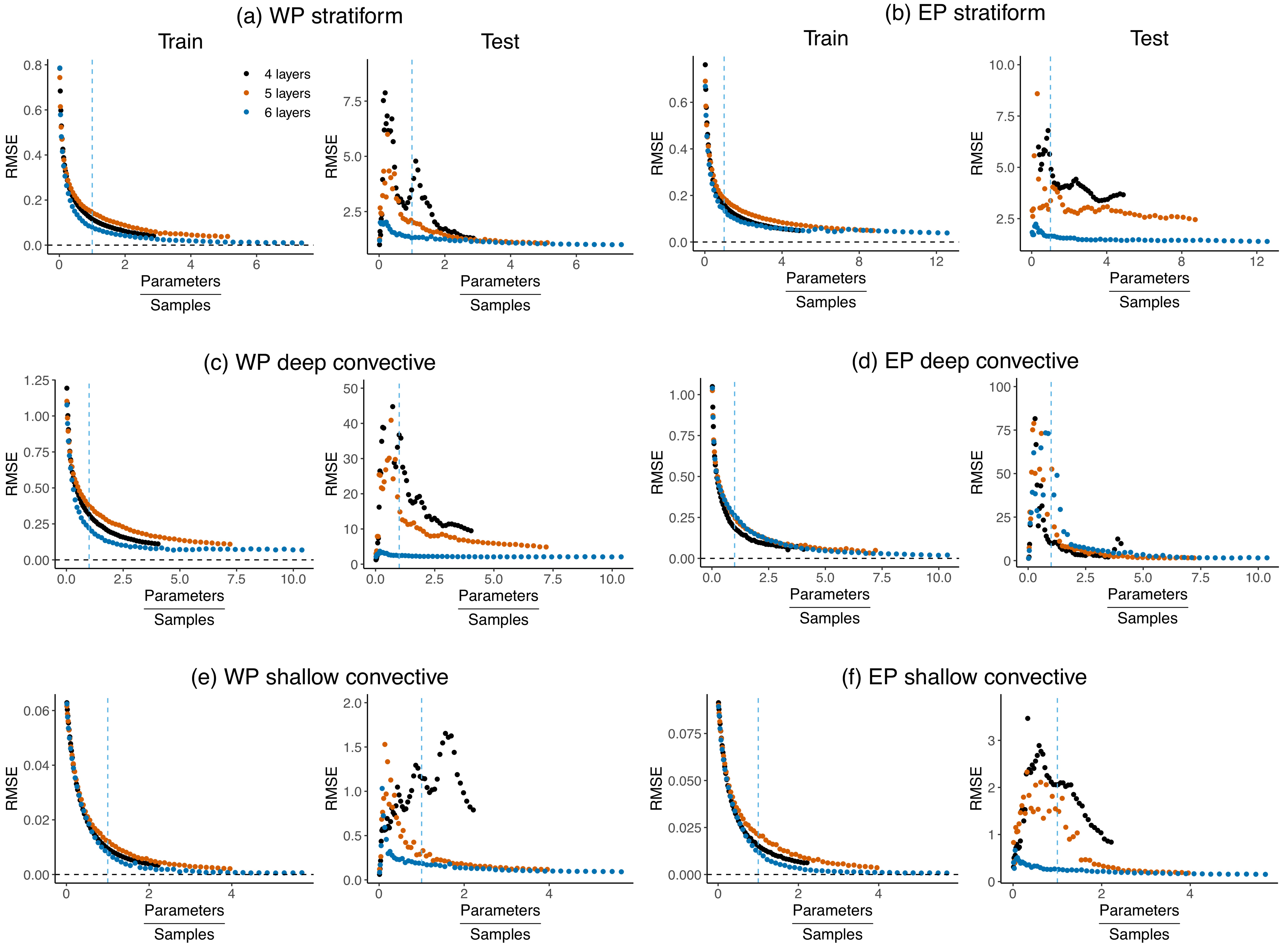}
    % \begin{subfigure}{0.49\textwidth}
    %     \subcaption{WP stratiform}
    %     \includegraphics[width=\textwidth]{WP-sf.pdf}
    % \end{subfigure}
    % \begin{subfigure}{0.49\textwidth}
    %     \subcaption{EP stratiform}
    %     \includegraphics[width=\textwidth]{EP-sf.pdf}
    % \end{subfigure}
    % \begin{subfigure}{0.49\textwidth}
    %     \subcaption{WP deep convective}
    %     \includegraphics[width=\textwidth]{WP-dp.pdf}
    % \end{subfigure}
    % \begin{subfigure}{0.49\textwidth}
    %     \subcaption{EP deep convective}
    %     \includegraphics[width=\textwidth]{EP-dp.pdf}
    % \end{subfigure}
    % \begin{subfigure}{0.49\textwidth}
    %     \subcaption{WP shallow convective}
    %     \includegraphics[width=\textwidth]{WP-sh.pdf}
    % \end{subfigure}
    % \begin{subfigure}{0.49\textwidth}
    %     \subcaption{EP shallow convective}
    %     \includegraphics[width=\textwidth]{EP-sh.pdf}
    % \end{subfigure}
    \caption{Training and test curves for the West Pacific (left two columns) and East Pacific (right two columns). The training results are the left panel and the test results are the right panel in each subfigure. The ratio of the number of parameters to the number of samples is on the x-axis, and the RMSE values are on the y-axis. The theoretical interpolation threshold is marked as a blue dashed line. Black, red, and blue curves represent NNs with 4, 5, and 6 layers, respectively.}
    \label{fig:descent}
\end{figure}

Figure \ref{fig:descent} illustrates the training and testing curves obtained from NNs with varying depths for different regions and rain types. As the model size increases, we are able to achieve nearly zero training error across all rain types. However, due to the heavy-tailed nature of the rainfall dataset, larger models were necessary to attain sufficiently low training error beyond the typical interpolation thresholds (i.e., where the sample size is equal to the number of model parameters as indicated by the vertical dashed line in each panel). In particular, NNs for deep convective rain need larger models to achieve nearly zero training error, as it is the most heavy-tailed rain type. \citet{nakkiran2021deep} also pointed out that other factors such as data distribution and training procedure may have influence on the model complexity, which is directly related to the double descent phenomenon. 

Notably, all the test curves in Figure \ref{fig:descent} exhibit the double descent phenomenon. The test errors after the peaks decrease and reach minimum or near-minimum points, although the peaks in the test curves occur at different positions. This observation holds true for all regions and rain types examined in the study, indicating that the over-parameterized NNs have been legitimately trained. A minor point to discuss is the position of the peaks in the test errors. Specifically, models with 6 layers often reach their peak before reaching the interpolation threshold.
In theory, with simple frameworks such as linear regression, the peaks are expected to occur at the interpolation threshold. However, a shift in the peak position has been observed in the literature and various reasons such as the data distribution, noises in responses, or regularization may cause the shift \citep{d2020double, nakkiran2021deep, gamba2022deep}. Henceforth, all over-parameterized NN results presented for each region and rain type are derived from the models at the rightmost end of the x-axis in Figure \ref{fig:descent}.

\subsection{Rain rate distributions}

\begin{figure}
    \centering
    \includegraphics[width=\linewidth]{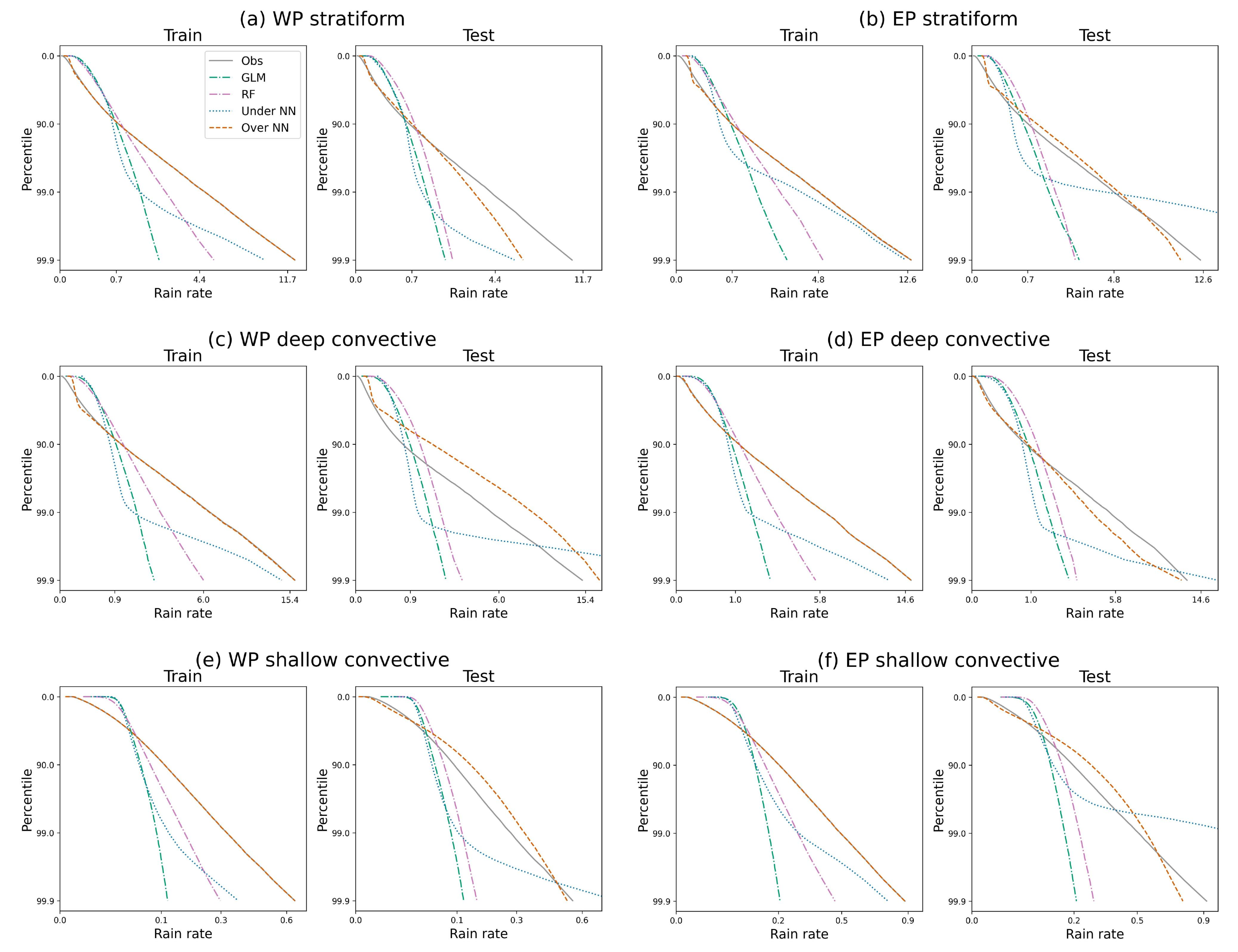}
    % \begin{subfigure}{0.49\textwidth}
    %     \subcaption{WP stratiform}
    %     \includegraphics[width=\textwidth]{WP-sf-full-percentile.jpg}
    % \end{subfigure}
    % \begin{subfigure}{0.49\textwidth}
    %     \subcaption{EP stratiform}
    %     \includegraphics[width=\textwidth]{EP-sf-full-percentile.jpg}
    % \end{subfigure}
    % \begin{subfigure}{0.49\textwidth}
    %     \subcaption{WP deep convective}
    %     \includegraphics[width=\textwidth]{WP-dp-full-percentile.jpg}
    % \end{subfigure}
    % \begin{subfigure}{0.49\textwidth}
    %     \subcaption{EP deep convective}
    %     \includegraphics[width=\textwidth]{EP-dp-full-percentile.jpg}
    % \end{subfigure}
    % \begin{subfigure}{0.49\textwidth}
    %     \subcaption{WP shallow convective}
    %     \includegraphics[width=\textwidth]{WP-sh-full-percentile.jpg}
    % \end{subfigure}
    % \begin{subfigure}{0.49\textwidth}
    %     \subcaption{EP shallow convective} 
    %     \includegraphics[width=\textwidth]{EP-sh-full-percentile.jpg}
    % \end{subfigure}
    \caption{Rain rate percentile plots (in mm/hr) for $0.5^{\circ}$ grids in the West Pacific and East Pacific regions for each rain type. Obs, Under NN, and Over NN stand for DPR observations, under-parameterized NN, and over-parameterized NN, respectively. Values on the x-axis indicate the 90\%, 99\% and 99.9\% percentiles from the training and test datasets for each domain and rain type.}
    \label{fig:percentile}
\end{figure}

Figure \ref{fig:percentile} illustrates percentile plots of estimated (Train) and predicted (Test) rain rate values from the GLM, RF, and NN methods over the WP and EP. Recall that the rain rates are based on $0.5^{\circ}$ grid means, so the values will be shifted lower compared to the native 5-km footprint resolution of the DPR. For both regions and all rain types, the distributions of estimated values from the over-parameterized NNs (red dashed line) successfully align with the distribution of the DPR observations (solid black line). This result is expected given that over-parameterized NNs achieved nearly zero training error in Figure \ref{fig:descent}. The other methods could not fully recover the training data and overestimate occurrence at low rain rates and underestimate occurrence at large rain rates. This result is consistent with \cite{yang2019predictive} and \cite{wang2021statistical}.

In the prediction results, over-parameterized NNs outperform the other methods and accurately capture the general shape of the rain rate distributions of the test dataset (Figure \ref{fig:percentile}). While the test lines for the over-parameterized NNs no longer lie directly on the observed DPR rain rates, they still hew closely to each other at both low to moderate percentiles (0 to 90\%) and tail values (90 to 99.9\%). %In fact, over-parameterized NNs achieve precise predictions of the rain rate distribution up to at least the 90th percentile rain rate for almost all rain types. In Figure \ref{fig:percentile}b, e and f, the prediction distributions almost match with the distribution of the test data up to 99.9th percentiles. This indicates over-parameterized NNs accomplishes very accurate predictions on the test set as well as almost perfect fits with the training set.
There are generally better rain rate predictions over the EP, similar to the GLM performance in \cite{yang2019predictive}. In particular, the predicted distribution of stratiform rain rate over the EP is very well represented from over-parameterized NNs. The deep convective prediction has especially high fidelity in the EP, whereas deep convective rain rates are all overpredicted in the WP. 

Figure \ref{fig:percentile} also shows that all the other methods exhibit markedly different patterns in their prediction distributions compared to the true rain rate distribution. Both GLM and RF tend to overpredict small rain rate values in the test data while underpredicting large values. This behavior is often seen in GCMs \citep[e.g.,][]{dai2006precipitation, fiedler2020simulated},  and indicates that these methods produce a narrower range of rain rate predictions compared to the wider distributions in over-parameterized NNs and the actual observations. Under-parameterized NNs initially show an underprediction at low rain rates, but then they bounce back to the observed rain rate distribution at higher values. However, under-parameterized NNs eventually end up generating too large of rain rates compared to the true values. Overall, these findings verify the superiority of the over-parameterized NNs in accurately predicting rain rate distributions.

Table \ref{tab:medae} provides quantitative comparison results between the different model rain rate distributions using median absolute errors (MedAE) \citep{longman2020characterizing, singh2021prediction, lucas2022optimizing, bai2023graph}, defined as follows:
$$\text{MedAE} = \text{median}(|y_{1}-\hat{y}_1|, \ldots, |y_n-\hat{y}_n|),$$
where $y_i$ and $\hat{y}_i$ ($i=1,\ldots,n$) represent the true observation and predicted value from a model, respectively, with a sample size of $n$. Despite the common use of RMSE and mean absolute error (MAE) \citep{wang2021statistical}, we identified their sensitivity to extreme values. Given the presence of a relatively significant proportion of extreme values in the rainfall dataset, and considering our focus on explaining those extreme values, a few erroneous yet extreme predictions could potentially distort the metrics. Consequently, methods that primarily produce small values might result in smaller RMSE or MAE values, which does not align with the objective of our study. In contrast, MedAE is renowned for its robustness to extreme values, making it a suitable choice for summarizing the outcomes of our study. In Table \ref{tab:medae}, across all cases, over-parameterized NNs consistently exhibit the smallest or closely ranked smallest MedAE values, suggesting that this method is producing more accurate predictions than the other candidates in general.

\begin{table}[!ht]
    \centering
    \caption{Median absolute errors (MedAE) of predicted values for each model, with values in bold indicating the smallest value among all methods considered.}
    \begin{tabular}{cccccc}\hline
        Region & Rain type & GLM & RF & Under NN & Over NN \\ 
        \hline\hline
        \multirow{3}{*}{WP} & Stratiform & 0.154 & 0.211 & 0.155 & {\bf 0.094} \\
         & Deep convective & 0.294 & 0.421 & 0.296 & {\bf 0.100} \\
         & Shallow convective & 0.031 & 0.036 & {\bf 0.030} & 0.036 \\ \hline
        \multirow{3}{*}{EP} & Stratiform & 0.174 & 0.241 & 0.150 & {\bf 0.103} \\
         & Deep Convective & 0.301 & 0.416 & 0.279 & {\bf 0.117} \\
         & Shallow Convective & 0.047 & 0.058 & {\bf 0.044} & 0.055 \\ \hline
    \end{tabular}
    \label{tab:medae}
\end{table}

\subsection{Spatial maps}

\begin{figure}[!ht]
    \centering
    {\small (a) Stratiform} \\
    \includegraphics[width=\textwidth]{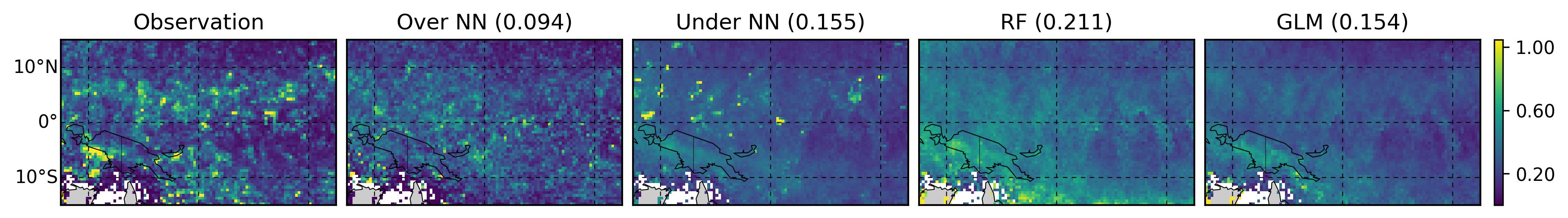}
    {\small (b) Deep Convective} \\
    \includegraphics[width=\textwidth]{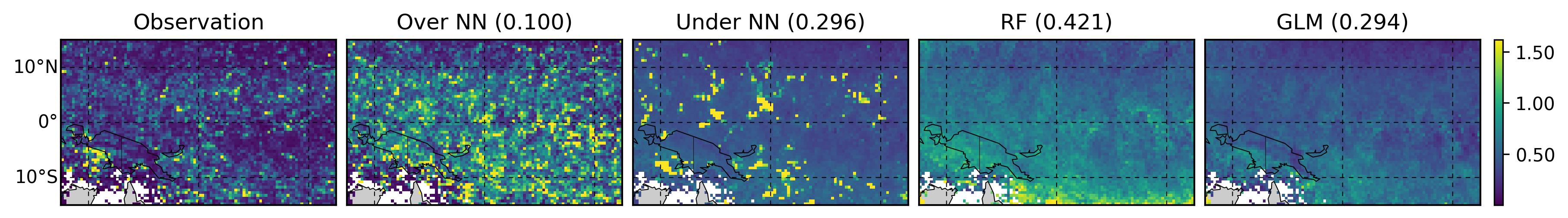}
    {\small (c) Shallow Convective} \\
    \includegraphics[width=\textwidth]{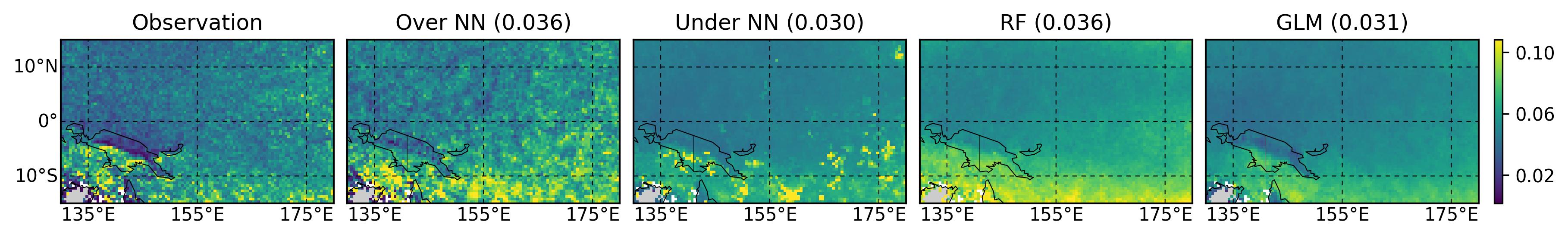}
    % \includegraphics[width=\textwidth]{WP_spatial_map_revised.pdf}
    % \begin{subfigure}{\textwidth}
    %     \subcaption{Stratiform}
    %     \includegraphics[width=\textwidth]{spatial-test-full-WP-sf-total.jpg}
    % \end{subfigure}
    % \begin{subfigure}{\textwidth}
    %     \subcaption{Deep convective}
    %     \includegraphics[width=\textwidth]{spatial-test-full-WP-dp-total.jpg}
    % \end{subfigure}
    % \begin{subfigure}{\textwidth}
    % \subcaption{Shallow convective}
    %     \includegraphics[width=\textwidth]{spatial-test-full-WP-sh-total.jpg}
    % \end{subfigure}
    \caption{Maps of the predicted mean $0.5^{\circ}$ JJA rain rates (in mm/hr) from 2019-2022 over the West Pacific region for (a) stratiform, (b) deep convective, and (c) shallow convective rain types using results from the over-parameterized NN, under-parameterized NN, RF, and GLM. 
    The medians of absolute error are presented in parentheses next to each model name.}
    \label{fig:WP_spatial_map}
\end{figure}

\begin{figure}[!ht]
    \centering
    {\small (a) Stratiform} \\
    \includegraphics[width=\textwidth]{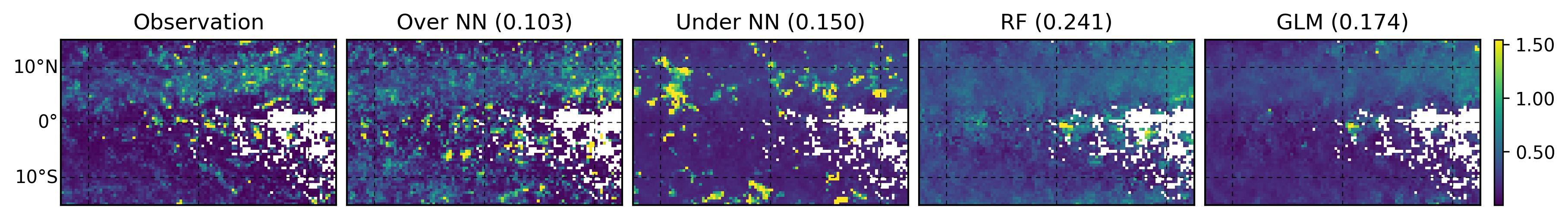}
    {\small (b) Deep Convective} \\
    \includegraphics[width=\textwidth]{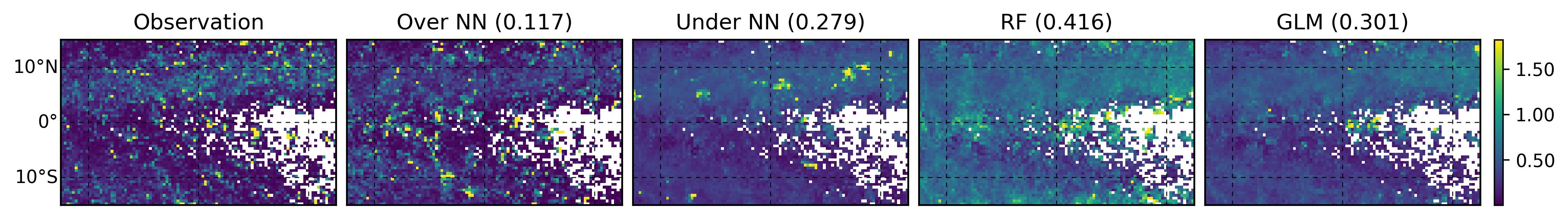}
    {\small (c) Shallow Convective} \\
    \includegraphics[width=\textwidth]{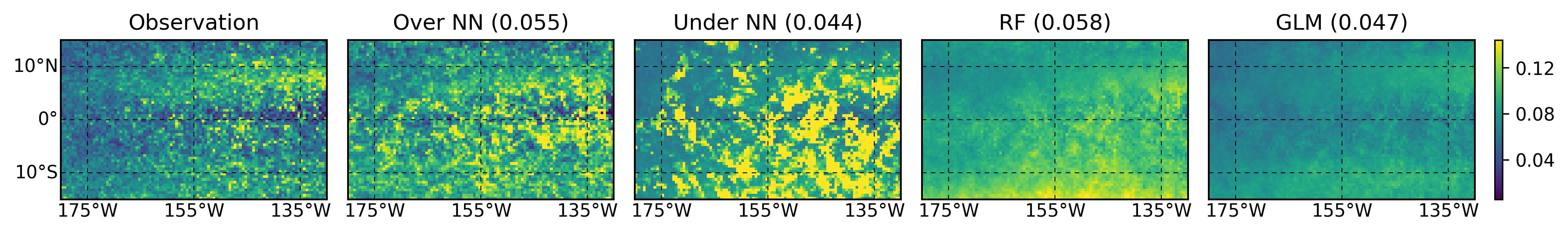}
    \caption{As in Figure \ref{fig:WP_spatial_map}, but over the East Pacific region.}
    \label{fig:EP_spatial_map}
\end{figure}

We next present the geographical patterns produced by each method. The rain rate prediction results are averaged over multiple years (2019-2022) to generate a single spatial map for each method. Figures \ref{fig:WP_spatial_map} and \ref{fig:EP_spatial_map} display averaged spatial maps over the WP and EP regions, respectively. While the observations indicate strong spatial heterogeneity over the domains for each rain type, RF and GLM produce very smooth prediction maps. Under-parameterized NNs produce non-smooth spatial patterns, but the high rain rates are too clustered and large. %In addition, the locations where the clusters appear have apparent distinctions from the observation spatial maps. 
Over-parameterized NNs, on the other hand, demonstrate their remarkable ability to capture the spatial heterogeneity of the DPR rain rate observations. Although the results do not portray individual weather events but rather a 4-year average, their pebble-like quality is much more representative of true weather events, compared to the under-parameterized NN boulders and ruffled sand of the RF and GLM maps. 

In terms of overall spatial patterns, all the methods capture the widespread rainfall of the WP warm pool (Figure \ref{fig:WP_spatial_map}) and the ITCZ in the northern part of the EP domain (Figure \ref{fig:EP_spatial_map}). However, the over-parameterized NN produces higher fidelity in the patterns. For example, while most of the training was done over ocean grid points, the over-parameterized NN provides more distinct predictions over Papua New Guinea in the southwest corner of the WP domain and is able to capture orographic features to a higher degree than the other methods, especially in the shallow convective field (Figure \ref{fig:WP_spatial_map}c). This result is promising for the future use of over-parameterized NNs for rainfall prediction over continents. In the EP, the over-parameterized NN produces a more distinct ITCZ in the stratiform and deep convective fields (Figures \ref{fig:EP_spatial_map}a and b), while also reproducing the pattern in shallow convective rainfall at the equator (Figure \ref{fig:EP_spatial_map}c) despite the fact that sea surface temperatures are not used as a predictor.

%In Figure \ref{fig:WP_spatial_map}c, the pattern of the observation map has heavy-rain area (yellow) and no-rain area (dark navy) next to each other at the bottom left. Over-parameterized NNs almost retrieved this pattern at the bottom left, while the other methods failed to do so. Similarly, in Figure \ref{fig:EP_spatial_map}a, the observation spatial maps show horizontal patterns in the area between $0-15^{\circ}$ and no-rain area for the rest of the map. Over-parameterized NNs accurately predicted the horizontal pattern in the same area and assigned small rain values to the no-rain area, aligning well with the observation spatial map. At last, the observation map in Figure \ref{fig:EP_spatial_map}-(c) exhibits symmetric horizontal yellow patterns around the equator. While the results from over-parameterized NNs are not perfect, they still manage to explain symmetric yellow patterns around the equator. These patterns do not show up in the prediction maps by the other methods where they lack either symmetry at all or clear yellow patterns. These results reinforce the outstanding capability of over-parameterized NNs in capturing spatial patterns in new data as well. 

\subsection{Interpreting over-parameterized neural networks}

In this section, we provide the feature importance for over-parameterized NNs with permutation importance. Table \ref{tab:perm} presents the 10 top features for the test output, categorized into five groups: humidity, temperature, zonal wind, meridional wind, and others. %The rationale behind grouping the features is that they carry meaning as categories rather than individual variables. 
Each category comprises values at 40 pressure levels except for the `Others' category, which contains fields reported at only one level (i.e., longitude, latitude, and latent heat flux). The individual feature contributions to the model output might not hold significant meaning, since permuting one variable while keeping the other variables the same may lead to unrealistic environmental conditions. Instead, our primary focus lies in identifying which category plays a significant role in the model output. Future work would be to analyze the individual and combined feature contributions in more detail.  
 
Higher numbers in Table \ref{tab:perm} represent a larger importance. For example, six pressure levels from the humidity category at or below 750 hPa are identified as the most significant for predicting shallow convective rain rates in the EP, while the other four most important features are temperature values at low (925 and 875 hPa) and mid-to-upper (400 and 350 hPa) levels. %The pressure levels associated with these profiles are 1000 mb, 975 mb, 950 mb, 925 mb, 775 mb, and 750 mb. 
While we used all available height levels (including the stratosphere) from MERRA-2 in our training and testing, we only report on the importance of pressure levels within the troposphere (i.e., up to 100 hPa), as almost all weather phenomena occur within this atmospheric layer. 

\begin{table}[!t]
    \centering
    \caption{
    Categories of the top 10 important features from over-parameterized NNs on test data for the West Pacific and East Pacific regions.
    %`Humidity', `Temperature', `Zonal wind', and `Meridional wind' represent 40 vertical profiles for each group. The `Others' category contains longitude, latitude, and latent heat flux. 
    The numbers in parentheses represent pressure levels (in hPa) of selected features. %Pressure levels within the troposphere, ranging from 1000 mb to 100 mb, are exclusively considered.
    }
    \begin{tabular}{ccccccccc}
    \topline
        Region & Rain type & Humidity & Temperature & Zonal wind & Meridional wind & Others \\ \midline 
        & &  & \\ [-1em]
        & Stratiform & 5 & 4  & 0 & 1  & 0 \\ [-.8em]
        & & {\scriptsize (875 800 775 725 500)} & {\scriptsize (1,000 925 800 650)} & & {\scriptsize (300)} & &\\ \cline{2-7}
        &  &  \\ [-1em]
        WP & Deep convective & 4 & 4 & 1 & 0 & 1 \\ [-.8em]
        & & {\scriptsize (875 700 550 450)} & {\scriptsize (900 800 775 700)} & {\scriptsize (550)} & & {\scriptsize (Latitude)} \\\cline{2-7}
        &  &  \\ [-1em]
        & Shallow convective & 5 & 3 & 0 & 2 & 0 \\ [-.8em]
        & & {\scriptsize (1,000 975 950 400 300)} & {\scriptsize (1,000 975 300)} &  & {\scriptsize (975 250)} \\\midline
         &  & \\ [-1em]
        & Stratiform & 2 & 5  & 2 & 0 & 1 \\ [-.8em]
        & & {\scriptsize (1,000 950)} & {\scriptsize (975 950 900 875 850)} & {\scriptsize (825 450)} & & {\scriptsize (Latitude)} \\ \cline{2-7}
        &  &  \\ [-1em]
        EP & Deep convective & 4 & 4 & 2 & 0 & 0 \\ [-.8em]
        & & {\scriptsize(725 700 600 350)} & {\scriptsize (925 850 400 175)} & {\scriptsize (125 100)} & &  \\\cline{2-7}
        &  &  \\ [-1em]
        & Shallow convective & 6 & 4 & 0 & 0 & 0 \\ [-.8em]
        & & {\scriptsize (1,000 975 950 925 775 750)} & {\scriptsize (925 875 400 350)} &  & \\ \hline
        \multicolumn{2}{c}{Average} & 4.33 & 4.00 & 0.83 & 0.50 & 0.33 \\ \botline
    \end{tabular}
    \label{tab:perm}
\end{table}

Table \ref{tab:perm} shows that the average number of variables in the humidity category that are selected as key features is 4.33, indicating that 4.33 variables from the humidity category were consistently chosen as the 10 most influential features in rain rate prediction across the tropical Pacific. This result is consistent with our common understanding that humidity has a direct relationship with rain rate \citep[e.g.,][]{bretherton2004relationships}, making it a critical factor in predicting rainfall. Furthermore, for most of the rain types, humidity contains the most variables among key features, which is consistent with the results of \cite{ahmed2015convective} that indicated a strong relationship between column humidity and shallow convective, deep convective, and stratiform rain rates across the tropical oceans. The humidity category is closely followed by temperature in importance, which aligns well with the results from \citet{yang2019predictive}. We note that most of the pressure levels highlighted for humidity and temperature are at or below 700 hPa. 

The most influential categories in explaining the rainfall prediction results over the tropical Pacific in Table \ref{tab:perm} after humidity and temperature are, in their respective order, zonal wind, meridional wind, and others. Even though wind (or wind shear) is not typically included as a parameter in GCM convective parameterizations, the small but perceptible presence of zonal and meridional wind in Table \ref{tab:perm} suggests that the inclusion of winds has the potential to improve convective and stratiform rain predictions. More precise quantification of these relationships are left for future studies, however, differences in the category rankings between the WP and EP and between rain types are consistent with our general understanding of large-scale environmental factors that affect rain production in the tropics.

\section{Discussion}

Capturing the observed climate variability in the tropical Pacific is one of the great challenges currently faced by climate modeling \citep{lee2022biases, sobel2023biases}. The rainfall response to sea surface temperature anomalies in this region plays a key role in driving this climate variability. In this study, we applied statistical and machine learning techniques to model the relationship between the large-scale environment and rainfall. With growing interest in the application of machine learning to climate modeling \citep{brenowitz2018prognostic, o2018using, rasp2018deep}, our results can help guide the development of new machine learning-derived parameterizations of rainfall.

We found that properly trained over-parameterized NNs correctly predicted the rain rate distributions over the tropical Pacific for multiple rain types (stratiform, deep convective, and shallow convective), including their tail behavior. Over-parameterized NNs also outperformed the other methods such as GLM and RF in both predicting the rain rate distribution and capturing spatial patterns. The permutation importance was implemented to address the feature importance in the NN model outcomes, producing consistent results with those obtained from the GLM framework in \cite{yang2019predictive}, who also found humidity and temperature to be the most important environmental variables. To the best of our knowledge, the benefits of over-parameterized models have been neither experimentally nor theoretically verified for heavy-tailed datasets. %Previous studies on over-parameterized NNs mainly focus on machine learning techniques and theoretical backgrounds within simplified frameworks. The performance of over-parameterized NNs has not been thoroughly explored with real-world datasets including heavy-tailed data. 
Given our successful results, the applicability of over-parameterized NNs can be expanded to various other real-world datasets. 

It is notable that the training and test curves in Figure \ref{fig:descent} show slightly different behaviors compared to previous studies. The interpolation thresholds have shifted to the right due to the heavy-tailed nature of the dataset, as we need larger model complexity to fully explain samples with large rain rate values. The locations of the peaks in the test curves were often placed much earlier than the interpolation thresholds. While it is partly known that factors like data distribution and noise in responses can influence the locations of the peaks, the majority of reasons and their impacts remain unclear and unexplored. One workaround with this issue is to consider $\ell_2$-regularization to the weight of NNs \citep{nakkiran2020optimal}. They found that with the optimally chosen tuning parameter, the peak can be completely eliminated, and even non-optimal regularization can alleviate the peak. Additionally, optimal early stopping based on test errors may also contribute to removing the peak, though this only happened for certain cases \citep{nakkiran2021deep}. Our study does not delve further into this matter, as our primary goal is to introduce over-parameterized NNs to rainfall prediction and examine the prediction performance of the method. Our findings indicate that no matter where the peaks are observed, we can attain minimum or near-minimum test errors in the over-parameterized regime and the obtained models correctly predict the rain rate distribution. We hope that future research, both in theory and applications, will shed light on the different patterns of the double descent curve that we observed with heavy-tailed data. 

In addition, the final models for each rain type and region have the potential for further improvement through fine-tuning and optimal selection of the hyperparameters. For instance, as observed in Figure \ref{fig:descent}, deeper neural networks consistently lead to better prediction performance even when the number of parameters is similar. The choice of activation functions and the optimizer also played a crucial role in determining the model performance. However, the process of searching for optimal hyperparameters needs to be done for each dataset individually, and it can be tedious work. Furthermore, the improvements achieved through hyperparameter tuning may not yield significant differences in the final results. We followed \citet{nakkiran2021deep} to choose the Adam optimizer over stochastic gradient descent and used an MLP structure to claim that we can obtain desirable results even with simple models. While further fine-tuning and customization of the model structure could potentially lead to improvements, the current method allows us to obtain promising results without the exhaustive search for hyperparameters. This helps maintain a balance between model performance and computational efficiency, making the approach more practical and applicable to real datasets. 

Lastly, we performed a number of sensitivity tests to see how flexible the over-parameterized NN framework is in predicting rain rates including i) not separating by rain type, ii) training on a combined WP/EP domain, and iii) considering both rain and no-rain events. The results (not shown) indicate that over-parameterized NNs still do better than GLM, RF, and under-parameterized NNs at capturing the tail and spatial structure of the rainfall over the tropical Pacific, but not as well as over-parameterized NNs that are trained for separate rain types or regions.
%pushed the boundaries of over-parameterized NN by applying them to the entire domain (WP and EP regions) with a single model and attempting to simultaneously predict both rain and no-rain events. While over-parameterized NNs over the entire domain successfully grasped the general rain rate trends, their predictions exhibited smoother patterns compared to real-world data or individually trained over-parameterized NN models. 
This highlights the limitations in capturing the full complexity of total rainfall over large regions. %which is inherently more intricate than individual regions. 
Furthermore, simultaneously modeling both rain and no-rain events proved challenging due to the dataset's zero-inflated and heavy-tailed nature. The training procedure encountered significant instability, highlighting the complexity of this task, so specialized architectures or optimization methods may be necessary. We leave this exploration for future research.

%\clearpage
%%%%%%%%%%%%%%%%%%%%%%%%%%%%%%%%%%%%%%%%%%%%%%%%%%%%%%%%%%%%%%%%%%%%%
% ACKNOWLEDGMENTS
%%%%%%%%%%%%%%%%%%%%%%%%%%%%%%%%%%%%%%%%%%%%%%%%%%%%%%%%%%%%%%%%%%%%%
\acknowledgments Aaron Funk processed the gridded radar and MERRA-2 data used in the analysis.
Mikyoung Jun acknowledges support by NSF IIS-2123247 and DMS-2105847. Courtney Schumacher acknowledges support by NASA 80NSSC22K0617. Funding for this project was also provided by the TAMU College of Arts \& Sciences Seed Funds.

%%%%%%%%%%%%%%%%%%%%%%%%%%%%%%%%%%%%%%%%%%%%%%%%%%%%%%%%%%%%%%%%%%%%%
% DATA AVAILABILITY STATEMENT
%%%%%%%%%%%%%%%%%%%%%%%%%%%%%%%%%%%%%%%%%%%%%%%%%%%%%%%%%%%%%%%%%%%%%
% 
%
\datastatement
The GPM DPR and MERRA-2 data are publicly available from the NASA GES DISC (https://disc.gsfc.nasa.gov/). 
The specific DPR and MERRA-2 training and test data sets used in this study have been placed on the Texas Data Repository (https://dataverse.tdl.org/).
Final over-parameterized models can be found in our Github repository: \href{https://github.com/HojunYou/Rainfall_prediction_with_overNN}{https://github.com/HojunYou/Rainfall\_prediction\_with\_overNN}.

%%%%%%%%%%%%%%%%%%%%%%%%%%%%%%%%%%%%%%%%%%%%%%%%%%%%%%%%%%%%%%%%%%%%%
% REFERENCES
%%%%%%%%%%%%%%%%%%%%%%%%%%%%%%%%%%%%%%%%%%%%%%%%%%%%%%%%%%%%%%%%%%%%%

\bibliographystyle{ametsocV6}
\bibliography{references}

\end{document}